\documentclass[]{article}
\pdfoutput=1
\usepackage{lmodern}
\usepackage{amssymb,amsmath}
\usepackage{ifxetex,ifluatex}
\usepackage{fixltx2e} 
\ifnum 0\ifxetex 1\fi\ifluatex 1\fi=0 
  \usepackage[T1]{fontenc}
  \usepackage[utf8]{inputenc}
\else 
  \ifxetex
    \usepackage{mathspec}
    \usepackage{xltxtra,xunicode}
  \else
    \usepackage{fontspec}
  \fi
  \defaultfontfeatures{Mapping=tex-text,Scale=MatchLowercase}
  
\fi
\IfFileExists{upquote.sty}{\usepackage{upquote}}{}
\IfFileExists{microtype.sty}{%
\usepackage{microtype}
\UseMicrotypeSet[protrusion]{basicmath} 
}{}
\usepackage[margin=1in]{geometry}
\usepackage{color}
\usepackage{fancyvrb}

\DefineVerbatimEnvironment{Highlighting}{Verbatim}{commandchars=\\\{\}}
\usepackage{framed}
\definecolor{shadecolor}{RGB}{248,248,248}
\newenvironment{Shaded}{\begin{snugshade}}{\end{snugshade}}
\newcommand{\KeywordTok}[1]{\textcolor[rgb]{0.13,0.29,0.53}{\textbf{{#1}}}}
\newcommand{\DataTypeTok}[1]{\textcolor[rgb]{0.13,0.29,0.53}{{#1}}}
\newcommand{\DecValTok}[1]{\textcolor[rgb]{0.00,0.00,0.81}{{#1}}}

\newcommand{\StringTok}[1]{\textcolor[rgb]{0.31,0.60,0.02}{{#1}}}
\newcommand{\CommentTok}[1]{\textcolor[rgb]{0.56,0.35,0.01}{\textit{{#1}}}}
\newcommand{\OtherTok}[1]{\textcolor[rgb]{0.56,0.35,0.01}{{#1}}}

\newcommand{\NormalTok}[1]{{#1}}
\usepackage{graphicx}
\makeatletter
\def\maxwidth{\ifdim\Gin@nat@width>\linewidth\linewidth\else\Gin@nat@width\fi}
\def\maxheight{\ifdim\Gin@nat@height>\textheight\textheight\else\Gin@nat@height\fi}
\makeatother
\setkeys{Gin}{width=\maxwidth,height=\maxheight,keepaspectratio}
\ifxetex
  \usepackage[setpagesize=false, 
              unicode=false, 
              xetex]{hyperref}
\else
  \usepackage[unicode=true]{hyperref}
\fi
\hypersetup{breaklinks=true,
            bookmarks=true,
            pdfauthor={Nicholas J. Horton, Benjamin S. Baumer, and Hadley Wickham},
            pdftitle={Setting the stage for data science: integration of data management skills in introductory and second courses in statistics},
            colorlinks=true,
            citecolor=blue,
            urlcolor=blue,
            linkcolor=magenta,
            pdfborder={0 0 0}}
\let\rmarkdownfootnote\footnote%
\def\footnote{\protect\rmarkdownfootnote}

\usepackage{titling}
  \title{Setting the stage for data science: integration of data management
skills in introductory and second courses in statistics}
  \pretitle{\vspace{\droptitle}\centering\huge}
  \posttitle{\par}
  \author{Nicholas J. Horton, Benjamin S. Baumer, and Hadley Wickham}
  \preauthor{\centering\large\emph}
  \postauthor{\par}
  \predate{\centering\large\emph}
  \postdate{\par}
  \date{February 1, 2015}

\begin{document}

\maketitle

Many have argued that statistics students need additional facility to
express statistical computations. By introducing students to commonplace
tools for data management, visualization, and reproducible analysis in
data science and applying these to real-world scenarios, we prepare them
to think statistically. In an era of increasingly big data, it is
imperative that students develop data-related capacities, beginning with
the introductory course. We believe that the integration of these
precursors to data science into our curricula---early and often---will
help statisticians be part of the dialogue regarding \emph{Big Data and
Big Questions}.

Specifically, through our shared experience working in industry,
government, private consulting, and academia we have identified five key
elements which deserve greater emphasis in the undergraduate curriculum
(in no particular order):

\begin{enumerate}
\itemsep1pt\parskip0pt\parsep0pt
\item
  Thinking creatively, but constructively, about data. This ``data
  tidying'' includes the ability to move data not only between different
  file formats, but also different \emph{shapes}. There are elements of
  data storage design (e.g.~normal forms) and foresight into how data
  should arranged based on how it will likely be used.
\item
  Facility with data sets of varying sizes and some understanding of
  scalability issues when working with data. This includes an elementary
  understanding of basic computer architecture (e.g.~memory vs.~hard
  disk space), and the ability to query a relational database management
  system (RDBMS).
\item
  Statistical computing skills in a command-driven environment (e.g.~R,
  Python, or Julia). Coding skills (in any language) are highly-valued
  and increasingly necessary. They provide freedom from the
  un-reproducible point-and-click application paradigm.
\item
  Experience wrestling with large, messy, complex, challenging data
  sets, for which there is no obvious goal or specially-curated
  statistical method (see SIDEBAR: What's in a name). While perhaps
  suboptimal for teaching specific statistical methods, these data are
  more similar to what analysts actually see in the wild.
\item
  An ethos of reproducibility. This is a major challenge for science in
  general, and we have the comparatively easy task of simply reproducing
  computations and analysis.
\end{enumerate}

We illustrate how these five elements can be addressed in the
undergraduate curriculum. To this end, we explore questions related to
airline travel using a large data set (point 4 above) that is by
necessity housed in a relational database (2, see SIDEBAR: Databases).
We present R code (3) using the \texttt{dplyr} framework (1) -- and
moreover, this paper itself -- in the reproducible R Markdown format
(5). Statistical educators play a key role in helping to prepare the
next generation of statisticians and data scientists. We hope that this
exercise will assist them in narrowing the aforementioned skills gap.

\paragraph{A framework for data-related
skills}\label{a-framework-for-data-related-skills}

The statistical data analysis cycle involves the formulation of
questions, collection of data, analysis, and interpretation of results
(see Figure 1). Data preparation and manipulation is not just a first
step, but a key component of this cycle (which will often be nonlinear,
see also \url{http://www.jstatsoft.org/v59/i10/paper}). When working
with data, analysts must first determine what is needed, describe this
solution in terms that a computer can understand, and execute the code.

\includegraphics{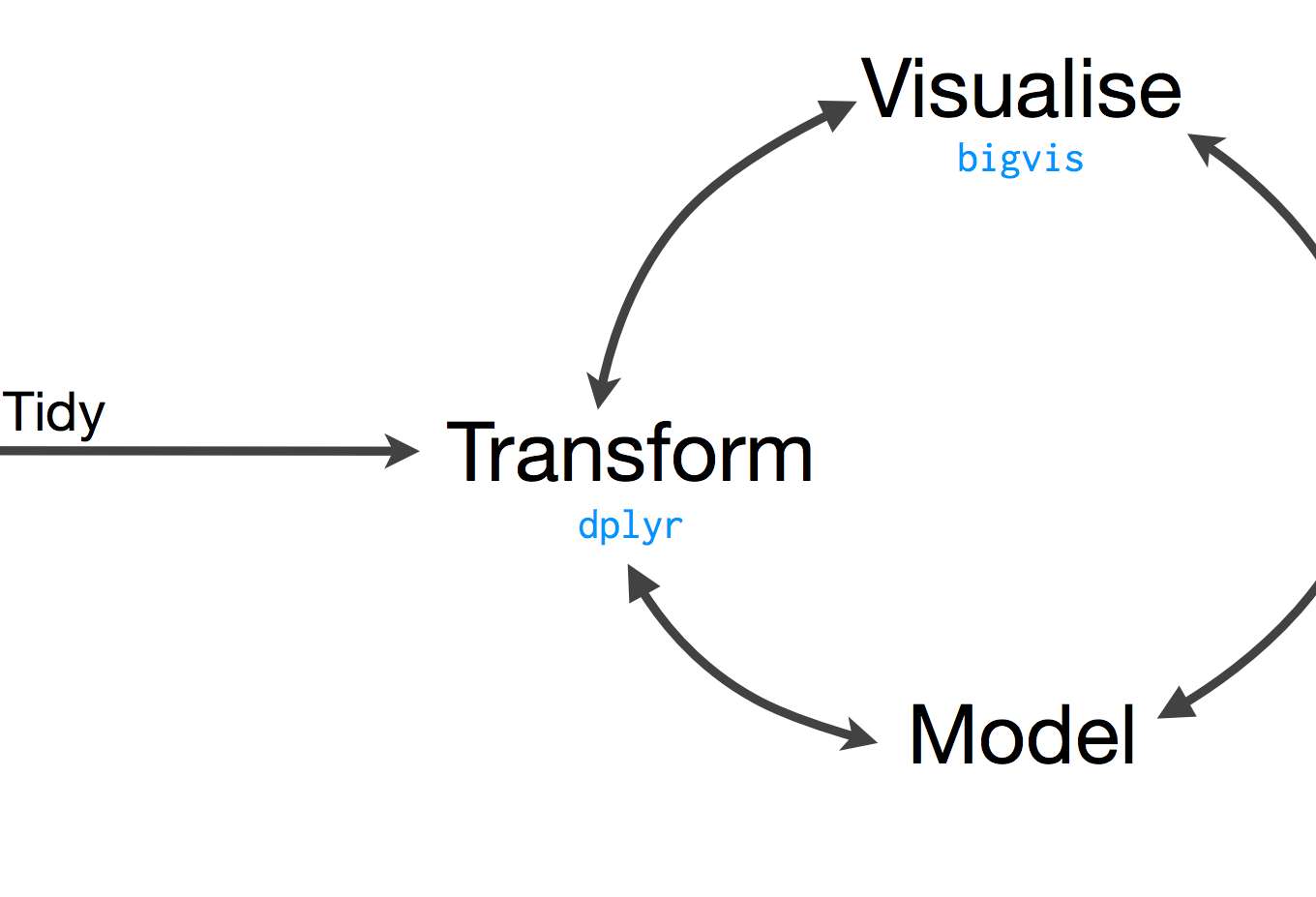} Figure 1: Statistical data analysis cycle
(source: \url{http://bit.ly/bigrdata4})

Here we illustrate how the \texttt{dplyr} package in R
(\url{http://cran.r-project.org/web/packages/dplyr}) can be used to
build a powerful and broadly accessible foundation for data
manipulation. This approach is attractive because it provides simple
functions that correspond to the most common data manipulation
operations (or \emph{verbs}) and uses efficient storage approaches so
that the analyst can focus on the analysis. (Other systems could
certainly be used in this manner, see for example
\url{http://iase-web.org/icots/9/proceedings/pdfs/ICOTS9_C134_CARVER.pdf}.)

\begin{verbatim}
verb          meaning
--------------------------------------------
select()      select variables (or columns)
filter()      subset observations (or rows)
mutate()      add new variables (or columns)
arrange()     re-order the observations
summarise()   reduce to a single row
group_by()    aggregate
left_join()   merge two data objects
distinct()    remove duplicate entries
collect()     force computation and bring data back into R
\end{verbatim}

Table 1: Key verbs in \texttt{dplyr} and \texttt{tidyr} to support data
management and manipulation (see \url{http://bit.ly/bigrdata4} for more
details)

\paragraph{Airline delays}\label{airline-delays}

To illustrate these data manipulation verbs in action, we consider
analysis of airline delays in the United States. This dataset,
constructed from information made available by the Bureau of
Transportation Statistics, was utilized in the ASA Data Expo 2009 (see
Wickham's paper in \emph{JCGS}). This rich data repository contains more
than 150 million observations corresponding to each commercial airline
flight in the United States between 1987 and 2012. (The magnitude of
this dataset precludes loading it directly in R, or most other general
purpose statistical packages.)

We demonstrate how to undertake analysis using the tools in the
\texttt{dplyr} package. (A smaller dataset is available for New York
City flights in 2013 within the \texttt{nycflights13} package. The
interface in R is almost identical in terms of the \texttt{dplyr}
functionality, with the same functions being used.)

Students can use this dataset to address questions that they find real
and relevant. (It is not hard to find motivation for investigating
patterns of flight delays. Ask students: have you ever been stuck in an
airport because your flight was delayed or cancelled and wondered if you
could have predicted the delay if you'd had more data?)

We begin by loading needed packages and connecting to a database
containing the flight, airline, airport, and airplane data (see SIDEBAR:
Databases).

\begin{Shaded}
\begin{Highlighting}[]
\KeywordTok{require}\NormalTok{(dplyr); }\KeywordTok{require}\NormalTok{(mosaic); }\KeywordTok{require}\NormalTok{(lubridate)    }\CommentTok{# login credentials in ~/.my.cnf}
\NormalTok{my_db <-}\StringTok{ }\KeywordTok{src_mysql}\NormalTok{(}\DataTypeTok{host=}\StringTok{"DBserver.com"}\NormalTok{, }\DataTypeTok{user=}\OtherTok{NULL}\NormalTok{, }\DataTypeTok{password=}\OtherTok{NULL}\NormalTok{, }\DataTypeTok{dbname=}\StringTok{"airlines"}\NormalTok{)}
\end{Highlighting}
\end{Shaded}

This example uses data from a database with multiple tables (collection
of related data).

\begin{Shaded}
\begin{Highlighting}[]
\NormalTok{ontime <-}\StringTok{ }\KeywordTok{tbl}\NormalTok{(my_db, }\StringTok{"ontime"}\NormalTok{)   }\CommentTok{# link to some useful tables}
\NormalTok{airports <-}\StringTok{ }\KeywordTok{tbl}\NormalTok{(my_db, }\StringTok{"airports"}\NormalTok{)}
\NormalTok{carriers <-}\StringTok{ }\KeywordTok{tbl}\NormalTok{(my_db, }\StringTok{"carriers"}\NormalTok{)}
\NormalTok{planes <-}\StringTok{ }\KeywordTok{tbl}\NormalTok{(my_db, }\StringTok{"planes"}\NormalTok{)}
\end{Highlighting}
\end{Shaded}

\paragraph{Filtering observations}\label{filtering-observations}

We start with an analysis focused on three smaller airports in the
Northeast. This illustrates the use of ${\tt filter()}$, which allows
the specification of a subset of rows of interest in the
\texttt{airports} table (or dataset). We first start by exploring the
\texttt{airports} table. Suppose we wanted to find out which airports
certain codes belong to?

\begin{Shaded}
\begin{Highlighting}[]
\KeywordTok{filter}\NormalTok{(airports, code %in%}\StringTok{ }\KeywordTok{c}\NormalTok{(}\StringTok{'ALB'}\NormalTok{, }\StringTok{'BDL'}\NormalTok{, }\StringTok{'BTV'}\NormalTok{))   }
\end{Highlighting}
\end{Shaded}

\begin{verbatim}
## Source: mysql 5.5.40-0ubuntu0.12.04.1 [DBuser@DBserver.com:/airlines]
## From: airports [3 x 7]
## Filter: code %in% c("ALB", "BDL", "BTV") 
## 
##   code                     name          city state country latitude
## 1  ALB               Albany Cty        Albany    NY     USA 42.74812
## 2  BDL    Bradley International Windsor Locks    CT     USA 41.93887
## 3  BTV Burlington International    Burlington    VT     USA 44.47300
## Variables not shown: longitude (dbl)
\end{verbatim}

\paragraph{Aggregating observations}\label{aggregating-observations}

Next we aggregate the counts of flights at all three of these airports
at the monthly level (in the \texttt{ontime} flight-level table), using
the \texttt{group\_by()} and \texttt{summarise()} functions. The
\texttt{collect()} function forces the evaluation. These functions are
connected using the \texttt{\%\textgreater{}\%} operator. This pipes the
results from one object or function as input to the next in an efficient
manner.

\begin{Shaded}
\begin{Highlighting}[]
\NormalTok{airportcounts <-}\StringTok{ }\NormalTok{ontime %>%}\StringTok{ }
\StringTok{   }\KeywordTok{filter}\NormalTok{(Dest %in%}\StringTok{ }\KeywordTok{c}\NormalTok{(}\StringTok{'ALB'}\NormalTok{, }\StringTok{'BDL'}\NormalTok{, }\StringTok{'BTV'}\NormalTok{)) %>%}
\StringTok{   }\KeywordTok{group_by}\NormalTok{(Year, Month, Dest) %>%}
\StringTok{   }\KeywordTok{summarise}\NormalTok{(}\DataTypeTok{count =} \KeywordTok{n}\NormalTok{()) %>%}\StringTok{ }
\StringTok{   }\KeywordTok{collect}\NormalTok{()}
\end{Highlighting}
\end{Shaded}

\paragraph{Creating new derived
variables}\label{creating-new-derived-variables}

Next we add a new column by constructing a date variable (using
\texttt{mutate()} and helper functions from the \texttt{lubridate}
package), then generate a time series plot.

\begin{Shaded}
\begin{Highlighting}[]
\NormalTok{airportcounts <-}\StringTok{ }\NormalTok{airportcounts %>%}
\StringTok{  }\KeywordTok{mutate}\NormalTok{(}\DataTypeTok{Date =} \KeywordTok{ymd}\NormalTok{(}\KeywordTok{paste}\NormalTok{(Year, }\StringTok{"-"}\NormalTok{, Month, }\StringTok{"-01"}\NormalTok{, }\DataTypeTok{sep=}\StringTok{""}\NormalTok{)))}
\KeywordTok{head}\NormalTok{(airportcounts) }\CommentTok{# list only the first six observations}
\end{Highlighting}
\end{Shaded}

\begin{verbatim}
## Source: local data frame [6 x 5]
## Groups: Year, Month
## 
##   Year Month Dest count       Date
## 1 1987    10  ALB   957 1987-10-01
## 2 1987    10  BDL  2580 1987-10-01
## 3 1987    10  BTV   549 1987-10-01
## 4 1987    11  ALB   950 1987-11-01
## 5 1987    11  BDL  2442 1987-11-01
## 6 1987    11  BTV   496 1987-11-01
\end{verbatim}

\begin{Shaded}
\begin{Highlighting}[]
\KeywordTok{xyplot}\NormalTok{(count ~}\StringTok{ }\NormalTok{Date, }\DataTypeTok{groups=}\NormalTok{Dest, }\DataTypeTok{type=}\KeywordTok{c}\NormalTok{(}\StringTok{"p"}\NormalTok{,}\StringTok{"l"}\NormalTok{), }\DataTypeTok{lwd=}\DecValTok{2}\NormalTok{, }\DataTypeTok{auto.key=}\KeywordTok{list}\NormalTok{(}\DataTypeTok{columns=}\DecValTok{3}\NormalTok{),}
       \DataTypeTok{ylab=}\StringTok{"Number of flights per month"}\NormalTok{, }\DataTypeTok{xlab=}\StringTok{"Year"}\NormalTok{, }\DataTypeTok{data=}\NormalTok{airportcounts)}
\end{Highlighting}
\end{Shaded}

\includegraphics{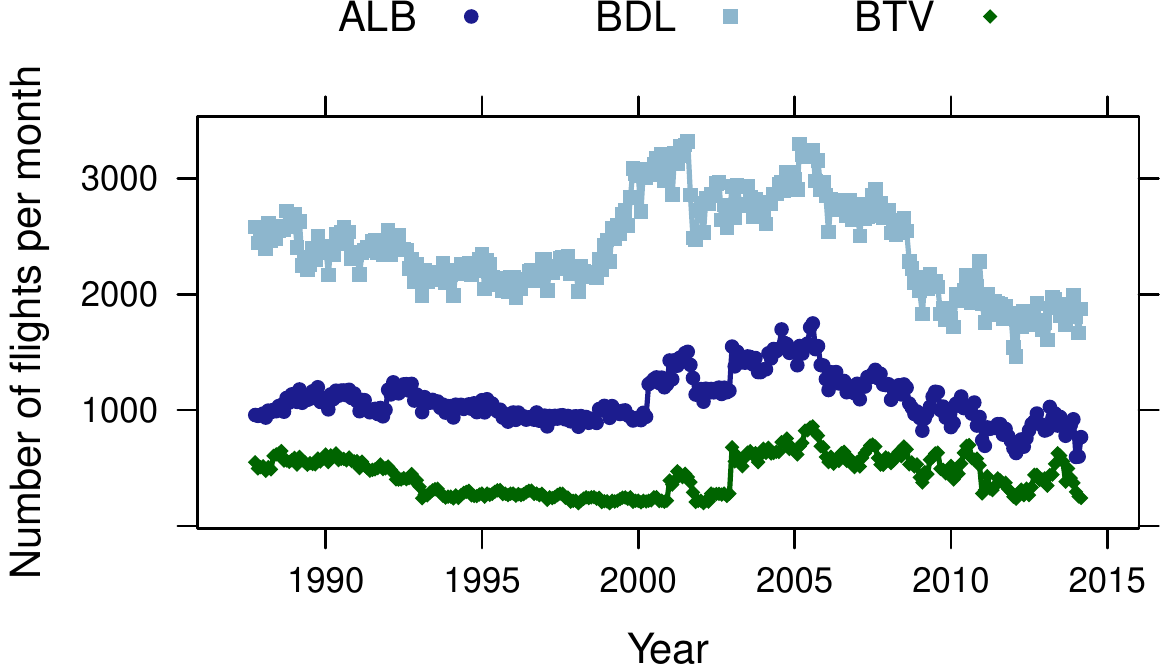}

Figure 2: Comparison of the number of flights arriving at three airports
over time

We observe in Figure 2 that there are some interesting patterns over
time for these airports. Bradley (serving Hartford, CT and Springfield,
MA) has the largest monthly volumes, with Burlington the least flights.
At all three airports, there has been a decline in the number of flights
from 2005 to 2012.

\paragraph{Sorting and selecting}\label{sorting-and-selecting}

Another important verb is \texttt{arrange()}, which in conjunction with
\texttt{head()} lets us display the months with the largest number of
flights. Here we need to use \texttt{ungroup()}, since otherwise the
data would remain aggregated by year, month, and destination.

\begin{Shaded}
\begin{Highlighting}[]
\NormalTok{airportcounts %>%}
\StringTok{  }\KeywordTok{ungroup}\NormalTok{() %>%}
\StringTok{  }\KeywordTok{arrange}\NormalTok{(}\KeywordTok{desc}\NormalTok{(count)) %>%}\StringTok{ }
\StringTok{  }\KeywordTok{select}\NormalTok{(count, Year, Month, Dest) %>%}\StringTok{ }
\StringTok{  }\KeywordTok{head}\NormalTok{() }
\end{Highlighting}
\end{Shaded}

\begin{verbatim}
## Source: local data frame [6 x 4]
## 
##   count Year Month Dest
## 1  3318 2001     8  BDL
## 2  3299 2005     3  BDL
## 3  3289 2001     7  BDL
## 4  3279 2001     5  BDL
## 5  3242 2005     8  BDL
## 6  3219 2005     5  BDL
\end{verbatim}

We can compare flight delays between two airlines serving a city pair.
For example, which airline was most reliable flying from Chicago O'Hare
(ORD) to Minneapolis/St.~Paul (MSP) in January, 2012? Here we
demonstrate how to calculate an average delay for each day for United,
Delta, and American (operated by Envoy/American Eagle). We create the
analytic dataset through use of ${\tt select()}$ (to pick the variables
to be included), \texttt{filter()} (to select a tiny subset of the
observations), and then repeat the previous aggregation. (Note that we
do not address flight cancellations: this exercise is left to the
reader, or see the online examples.)

\begin{Shaded}
\begin{Highlighting}[]
\NormalTok{delays <-}\StringTok{ }\NormalTok{ontime %>%}\StringTok{ }
\StringTok{  }\KeywordTok{select}\NormalTok{(Origin, Dest, Year, Month, DayofMonth, UniqueCarrier, ArrDelay) %>%}
\StringTok{  }\KeywordTok{filter}\NormalTok{(Origin ==}\StringTok{ 'ORD'} \NormalTok{&}\StringTok{ }\NormalTok{Dest ==}\StringTok{ 'MSP'} \NormalTok{&}\StringTok{ }\NormalTok{Year ==}\StringTok{ }\DecValTok{2012} \NormalTok{&}\StringTok{ }\NormalTok{Month ==}\StringTok{ }\DecValTok{1} \NormalTok{&}\StringTok{ }
\StringTok{         }\NormalTok{(UniqueCarrier %in%}\StringTok{ }\KeywordTok{c}\NormalTok{(}\StringTok{"UA"}\NormalTok{, }\StringTok{"MQ"}\NormalTok{, }\StringTok{"DL"}\NormalTok{))) %>%}
\StringTok{  }\KeywordTok{group_by}\NormalTok{(Year, Month, DayofMonth, UniqueCarrier) %>%}
\StringTok{  }\KeywordTok{summarise}\NormalTok{(}\DataTypeTok{meandelay =} \KeywordTok{mean}\NormalTok{(ArrDelay), }\DataTypeTok{count =} \KeywordTok{n}\NormalTok{()) %>%}
\StringTok{  }\KeywordTok{collect}\NormalTok{()}
\end{Highlighting}
\end{Shaded}

\paragraph{Merging}\label{merging}

Merging is another key capacity for students to master. Here, the full
carrier names are merged (or joined, in database parlance) to facilitate
the comparison, using the \texttt{left\_join()} function to provide a
less terse full name for the airlines in the legend of the figure.

\begin{Shaded}
\begin{Highlighting}[]
\NormalTok{carriernames <-}\StringTok{ }\NormalTok{carriers %>%}
\StringTok{  }\KeywordTok{filter}\NormalTok{(code %in%}\StringTok{ }\KeywordTok{c}\NormalTok{(}\StringTok{"UA"}\NormalTok{, }\StringTok{"MQ"}\NormalTok{, }\StringTok{"DL"}\NormalTok{)) %>%}
\StringTok{  }\KeywordTok{collect}\NormalTok{()}
\NormalTok{merged <-}\StringTok{ }\KeywordTok{left_join}\NormalTok{(delays, carriernames, }\DataTypeTok{by=}\KeywordTok{c}\NormalTok{(}\StringTok{"UniqueCarrier"} \NormalTok{=}\StringTok{ "code"}\NormalTok{))}
\KeywordTok{head}\NormalTok{(merged) }
\end{Highlighting}
\end{Shaded}

\begin{verbatim}
## Source: local data frame [6 x 7]
## Groups: Year, Month, DayofMonth
## 
##   Year Month DayofMonth UniqueCarrier meandelay count
## 1 2012     1          1            MQ    8.7500     4
## 2 2012     1          1            UA   -5.0000     2
## 3 2012     1          2            DL    2.3333     3
## 4 2012     1          2            MQ   24.1667     6
## 5 2012     1          2            UA   -3.0000     3
## 6 2012     1          3            DL    8.0000     2
## Variables not shown: name (chr)
\end{verbatim}

\begin{Shaded}
\begin{Highlighting}[]
\KeywordTok{densityplot}\NormalTok{(~}\StringTok{ }\NormalTok{meandelay, }\DataTypeTok{group=}\NormalTok{name, }\DataTypeTok{auto.key=}\OtherTok{TRUE}\NormalTok{, }\DataTypeTok{xlab=}\StringTok{"Average daily delay (in minutes)"}\NormalTok{, }
  \DataTypeTok{data=}\NormalTok{merged)}
\end{Highlighting}
\end{Shaded}

\includegraphics{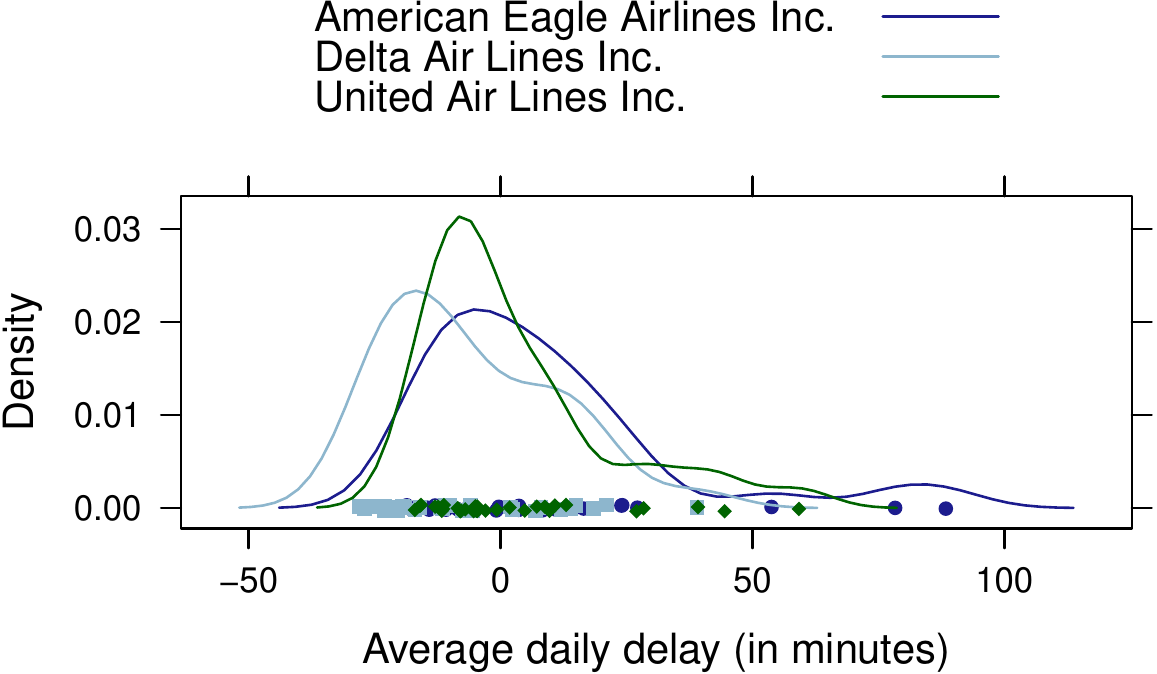}

Figure 3: Comparison of mean flight delays from O'Hare to
Minneapolis/St.~Paul in January, 2012 for three airlines

We see in Figure 3 that the airlines are fairly reliable, though there
were some days with average delays of 50 minutes or more (three of which
were accounted for by Envoy/American Eagle).

\begin{Shaded}
\begin{Highlighting}[]
\KeywordTok{filter}\NormalTok{(delays, meandelay >}\StringTok{ }\DecValTok{50}\NormalTok{)}
\end{Highlighting}
\end{Shaded}

\begin{verbatim}
## Source: local data frame [4 x 6]
## Groups: Year, Month, DayofMonth
## 
##   Year Month DayofMonth UniqueCarrier meandelay count
## 1 2012     1         13            MQ   78.3333     6
## 2 2012     1         20            UA   59.2500     6
## 3 2012     1         21            MQ   88.4000     5
## 4 2012     1         22            MQ   53.8000     6
\end{verbatim}

Finally, we can drill down to explore all flights on Mondays in the year
2001.

\begin{Shaded}
\begin{Highlighting}[]
\NormalTok{flights <-}\StringTok{ }\NormalTok{ontime %>%}
\StringTok{  }\KeywordTok{filter}\NormalTok{(DayofWeek==}\DecValTok{1} \NormalTok{&}\StringTok{ }\NormalTok{Year==}\DecValTok{2001}\NormalTok{) %>%}\StringTok{ }
\StringTok{  }\KeywordTok{group_by}\NormalTok{(Year, Month, DayOfMonth) %>%}
\StringTok{  }\KeywordTok{summarise}\NormalTok{(}\DataTypeTok{count =} \KeywordTok{n}\NormalTok{()) %>%}\StringTok{ }
\StringTok{  }\KeywordTok{collect}\NormalTok{()}
\NormalTok{flights <-}\StringTok{ }\KeywordTok{mutate}\NormalTok{(flights,}\DataTypeTok{Date=}\KeywordTok{ymd}\NormalTok{(}\KeywordTok{paste}\NormalTok{(Year, }\StringTok{"-"}\NormalTok{, Month, }\StringTok{"-"}\NormalTok{, DayOfMonth, }\DataTypeTok{sep=}\StringTok{""}\NormalTok{)))}
\KeywordTok{xyplot}\NormalTok{(count ~}\StringTok{ }\NormalTok{Date, }\DataTypeTok{type=}\StringTok{"l"}\NormalTok{, }\DataTypeTok{ylab=}\StringTok{"Count of flights on Mondays"}\NormalTok{, }\DataTypeTok{data=}\NormalTok{flights)}
\KeywordTok{ladd}\NormalTok{(}\KeywordTok{panel.abline}\NormalTok{(}\DataTypeTok{v=}\KeywordTok{ymd}\NormalTok{(}\StringTok{"2001-09-11"}\NormalTok{), }\DataTypeTok{lty=}\DecValTok{2}\NormalTok{))}
\end{Highlighting}
\end{Shaded}

\includegraphics{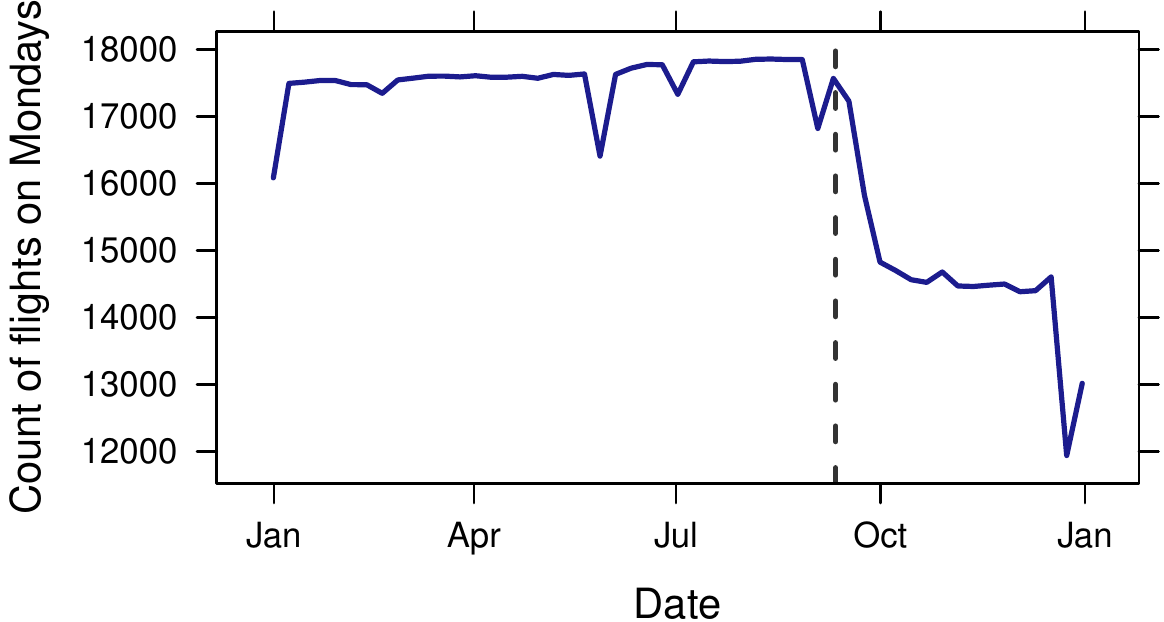}

Figure 4: display of counts of commercial flights on Mondays in the
United States in 2001. The clear impact of 9/11 can be seen.

\paragraph{Integrating bigger questions and datasets into the
curriculum}\label{integrating-bigger-questions-and-datasets-into-the-curriculum}

This opportunity to make a complex and interesting dataset accessible to
students in introductory statistics is quite compelling. In the
introductory (or first) statistics course, we explored airline delays
without any technology through use of the ``Judging Airlines'' model
eliciting activity (MEA) developed by the CATALST Group
(\url{http://serc.carleton.edu/sp/library/mea/examples/example5.html}).
This MEA guides students to develop ideas regarding center and
variability and the basics of informal inferences using small samples of
data for pairs of airlines flying out of Chicago.

\includegraphics{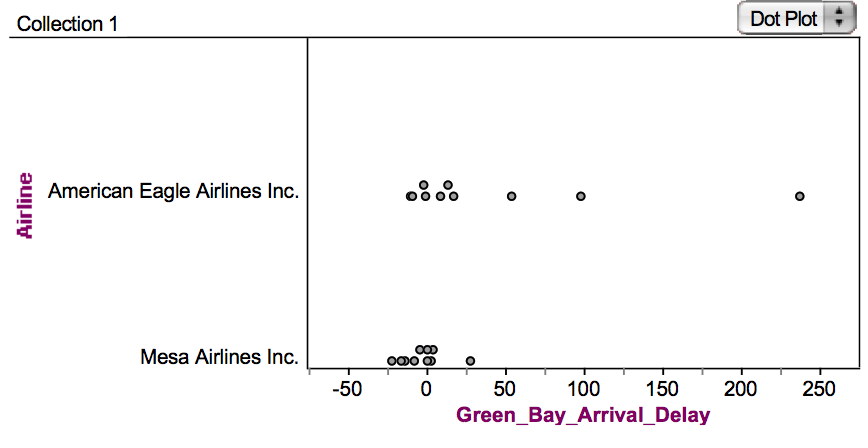}

Figure 5: Fathom software display of sample airline delays data for a
city pair used in the ``Judging Airlines'' MEA (model eliciting
activity)

Figure 5 displays sample airline delays for ten flights each for
American Eagle Airlines and Mesa Airlines flying from Chicago to Green
Bay, Wisconsin. As part of this activity, students need to describe five
possible sample statistics which could be used to compare the flight
delays by airline. These might include the average, the maximum, the
median, the 90th percentile, or the fraction that are late. Finally,
they need to create a rule that incorporates at least two of those
summary statistics that can be used to make a decision about whether one
airline is more reliable. A possible rule might be to declare an airline
is better than another if that airline has half an hour less average
delay, and that same airline has 10\% less delayed flights than the
other (if the two measures of reliability differ in direction for the
two airlines, no call is made).

To finish the assignment, students are provided with data for another
four city pairs, asked to carry out their rule on these new ``test''
datasets, then summarize their results in a letter to the editor of
\emph{Chicago Magazine}.

Later in the course, the larger dataset can be reintroduced in several
ways. It can be brought into class to illustrate univariate summaries or
bivariate relationships (including more sophisticated visualization and
graphical displays). Students can pose questions through projects or
other extended assignments. A lab activity could have students explore
their favorite airport or city pair (when comparing two airlines they
will often find that only one airline services that connection,
particularly for smaller airports.) Students could be asked to return to
the informal ``rule'' they developed in an extension to assess its
performance. Their rule can be programmed in R, and then carried out on
a series of random samples from the flights from that city on that
airline within that year. This allows them to see how often their rule
picked an airline as being more reliable (using various subsets of the
observed data as the ``truth''). Finally, students can summarize the
population of all flights, as a way to better understand sampling
variability. This process reflects the process followed by analysts
working with big data: sampling is used to generate hypotheses that are
then tested against the complete dataset.

In a second course, more time is available to develop diverse
statistical and computational skills. This includes more sophisticated
data management and manipulation with explicit learning outcomes that
are a central part of the course syllabus.

Other data wrangling and manipulation capacities can be introduced and
developed using this example, including more elaborate data joins/merges
(since there are tables providing additional (meta)data about planes).
As an example, consider the many flights of plane N355NB, which flew out
of Bradley airport in January, 2008.

\begin{Shaded}
\begin{Highlighting}[]
\KeywordTok{filter}\NormalTok{(planes, tailnum==}\StringTok{"N355NB"}\NormalTok{)}
\end{Highlighting}
\end{Shaded}

\begin{verbatim}
## Source: mysql 5.5.40-0ubuntu0.12.04.1 [DBuser@DBserver.com:/airlines]
## From: planes [1 x 10]
## Filter: tailnum == "N355NB" 
## 
##   tailnum        type manufacturer issue_date    model status
## 1  N355NB Corporation       AIRBUS 11/14/2002 A319-114  Valid
## Variables not shown: aircraft_type (chr), engine_type (chr), year (int),
##   issueDate (chr)
\end{verbatim}

We see that this is an Airbus 319.

\begin{Shaded}
\begin{Highlighting}[]
\NormalTok{singleplane <-}\StringTok{ }\KeywordTok{filter}\NormalTok{(ontime, tailnum==}\StringTok{"N355NB"}\NormalTok{) %>%}\StringTok{ }
\StringTok{  }\KeywordTok{select}\NormalTok{(Year, Month, DayofMonth, Dest, Origin, Distance) %>%}
\StringTok{  }\KeywordTok{collect}\NormalTok{()}
\NormalTok{singleplane %>%}\StringTok{ }
\StringTok{  }\KeywordTok{group_by}\NormalTok{(Year) %>%}
\StringTok{  }\KeywordTok{summarise}\NormalTok{(}\DataTypeTok{count =} \KeywordTok{n}\NormalTok{(), }\DataTypeTok{totaldist =} \KeywordTok{sum}\NormalTok{(Distance))}
\end{Highlighting}
\end{Shaded}

\begin{verbatim}
## Source: local data frame [13 x 3]
## 
##    Year count totaldist
## 1  2002   152    136506
## 2  2003  1367   1224746
## 3  2004  1299   1144288
## 4  2005  1366   1149142
## 5  2006  1484   1149036
## 6  2007  1282   1010146
## 7  2008  1318   1095109
## 8  2009  1235   1094532
## 9  2010  1368   1143189
## 10 2011  1406    919893
## 11 2012  1213    807246
## 12 2013  1339    921673
## 13 2014   169     97257
\end{verbatim}

\begin{Shaded}
\begin{Highlighting}[]
\KeywordTok{sum}\NormalTok{(~}\StringTok{ }\NormalTok{Distance, }\DataTypeTok{data=}\NormalTok{singleplane)}
\end{Highlighting}
\end{Shaded}

\begin{verbatim}
## [1] 11892763
\end{verbatim}

We see that this Airbus A319 has been very active, with around 1300
flights per year since it came online in late 2002 and has amassed more
than 11 million miles in the air.

\begin{Shaded}
\begin{Highlighting}[]
\NormalTok{singleplane %>%}
\StringTok{  }\KeywordTok{group_by}\NormalTok{(Dest) %>%}
\StringTok{  }\KeywordTok{summarise}\NormalTok{(}\DataTypeTok{count =} \KeywordTok{n}\NormalTok{()) %>%}\StringTok{ }
\StringTok{  }\KeywordTok{arrange}\NormalTok{(}\KeywordTok{desc}\NormalTok{(count)) %>%}\StringTok{ }
\StringTok{  }\KeywordTok{filter}\NormalTok{(count >}\StringTok{ }\DecValTok{500}\NormalTok{)}
\end{Highlighting}
\end{Shaded}

\begin{verbatim}
## Source: local data frame [5 x 2]
## 
##   Dest count
## 1  MSP  2861
## 2  DTW  2271
## 3  MEM   847
## 4  LGA   812
## 5  ATL   716
\end{verbatim}

Finally, we see that it tends to spend much of its time flying to
Minneapolis/St.~Paul (MSP) and Detroit (DTW).

\paragraph{Mapping}\label{mapping}

Mapping is also possible, since the latitude and longitude of the
airports are provided. Figure 6 displays a map of flights from Bradley
airport in 2013 (the code to create the display can be found with the
online examples).

\includegraphics{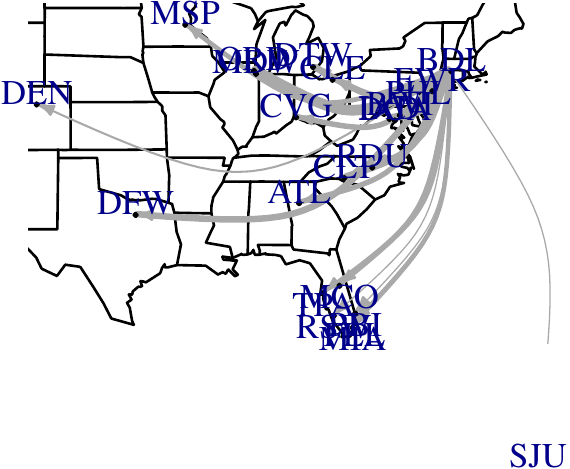}

Figure 6: Map of flights from Bradley airport in 2013

Linkage to other data scraped from the Internet (e.g.~detailed weather
information for a particular airport or details about individual planes)
may allow other questions to be answered (this has already been included
in the \texttt{nycflights13} package (see SIDEBAR: Databases). Use of
this rich dataset helps to excite students about the power of
statistics, introduce tools that can help energize the next generation
of data scientists, and build useful data-related skills.

\subsubsection{Conclusion and next
steps}\label{conclusion-and-next-steps}

Statistics students need to develop the capacity to make sense of the
staggering amount of information collected in our increasingly
data-centered world. In her 2013 book, Rachel Schutt succinctly
summarized the challenges she faced as she moved into the workforce:
``It was clear to me pretty quickly that the stuff I was working on at
Google was different than anything I had learned at school.'' This
anecdotal evidence is corroborated by the widely cited McKinsey report
that called for the training of hundreds of thousands of workers with
the skills to make sense of the rich and sophisticated data now
available to make decisions (along with millions of new managers with
the ability to comprehend these results). The disconnect between the
complex analyses now demanded in industry and the instruction available
in academia is a major challenge for the profession.

We agree that there are barriers and time costs to the introduction of
reproducible analysis tools and more sophisticated data management and
manipulation skills to our courses. Further guidance and research
results are needed to guide our work in this area, along with
illustrated examples, case studies, and faculty development. But these
impediments must not slow down our adoption. As Schutt cautions in her
book, statistics could be viewed as obsolete if this challenge is not
embraced. We believe that the time to move forward in this manner is
now, and believe that these these basic data-related skills provide a
foundation for such efforts.

Copies of the R Markdown and formatted files for these analyses (to
allow replication of the analyses) along with further background on
databases and the Airline Delays dataset are available at
\url{http://www.amherst.edu/~nhorton/precursors}. A previous version of
this paper was presented in July, 2014 at the International Conference
on Teaching Statistics (ICOTS9) in Flagstaff, AZ.  Partial support for this work was
made available by NSF grant 0920350 (Project MOSAIC, \url{http://www.mosaic-web.org}).

\paragraph{Further reading}\label{further-reading}

American Statistical Association Undergraduate Guidelines Workgroup
(2014). 2014 Curriculum guidelines for undergraduate programs in
statistical science. Alexandria, VA: American Statistical Association,
\url{http://www.amstat.org/education/curriculumguidelines.cfm}.

Baumer, B., Cetinkaya-Rundel, M., Bray, A., Loi, L. and Horton, N.J.
(2014). R Markdown: Integrating a reproducible analysis tool into
introductory statistics, \emph{Technology Innovations in Statistics
Education}, \url{http://escholarship.org/uc/item/90b2f5xh}.

Finzer, W. (2013). The data science education dilemma. \emph{Technology
Innovations in Statistics Education},
\url{http://escholarship.org/uc/item/7gv0q9dc}.

Horton, N.J., Baumer, Ben S., and Wickham, H. (2014). Teaching
precursors to data science in introductory and second courses in
statistics, \url{http://arxiv.org/abs/1401.3269}.

Nolan, D. and Temple Lang, D. (2010). Computing in the statistics
curricula, \emph{The American Statistician}, 64, 97--107.

O'Neil, C. and Schutt R. (2013). Doing Data Science: Straight Talk from
the Frontline, O'Reilly and Associates.

Wickham, H. (2011). ASA 2009 Data Expo, \emph{Journal of Computational
and Graphical Statistics}, 20(2):281-283.

\subsubsection{SIDEBAR: What's in a
word?}\label{sidebar-whats-in-a-word}

In their 2010 American Statistician paper, Deborah Nolan and Duncan
Temple Lang describe the need for students to be able to ``compute with
data'' to be able to answer statistical questions. Diane Lambert of
Google calls this the capacity to ``think with data''. Statistics
graduates need to be manage data, analyze it accurately, and communicate
findings effectively. The Wikipedia data science entry states that
``data scientists use the ability to find and interpret rich data
sources, manage large amounts of data despite hardware, software, and
bandwidth constraints, merge data sources, ensure consistency of
datasets, create visualizations to aid in understanding data, build
mathematical models using the data, present and communicate the data
insights/findings to specialists and scientists in their team and if
required to a non-expert audience.'' But what is the best word or phrase
to describe these computational and data-related skills?

``Data wrangling'' has been suggested as one possibility (and returned
about 131,000 results on Google), though this connotes the idea of a
long and complicated dispute, often involving livestock, which may not
end well.

``Data grappling'' is another option (about 7,500 results on Google),
though this perhaps less attractive as it suggests pirates (and
grappling hooks) or wrestling as combat sport or self defense.

``Data munging'' (about 35,000 results on Google) is a common term in
computer science used to describe changes to data (both constructive and
destructive) or mapping from one format to another. A disadvantage of
this term is that it has a somewhat pejorative sentiment.

``Data tidying'' (about 900 results on Google) brings to mind the ideas
of ``bringing order to'' or ``arranging neatly''.

``Data curation'' (about 322,000 results on Google) is a term that
focuses on a long-term time scale for use (and preservation). While
important, this may be perceived a dusty and stale task.

``Data cleaning'' (or ``data cleansing'', about 490,000 results on
Google) is the process to identify and correct (or remove) invalid
records from a dataset. Other related terms include ``data
standardization'' and ``data harmonization''.

A search for ``Data manipulation'' yielded about 740,000 results on
Google. Interestingly, this term on Wikipedia redirects to the ``Misuse
of statistics'' page, implying the analyst might have malicious
intentions and could torture the data to tell a particular story. The
Wikipedia ``Data manipulation language'' page has no such negative
connotations (and describes the Structured Query Language {[}SQL{]} as
one such language). This dual meaning stems from the definition (from
Merriam-Webster) of manipulate:

\begin{itemize}
\itemsep1pt\parskip0pt\parsep0pt
\item
  To manage or utilize skillfully
\item
  To control or play upon by artful, unfair, or insidious means
  especially to one's own advantage
\end{itemize}

``Data management'' was the most common term, with more than 33,000,000
results on Google. The DAMA Data Management Body of Knowledge
(DAMA-DMBOK,
\url{http://www.dama.org/files/public/DI_DAMA_DMBOK_Guide_Presentation_2007.pdf})
provides a definition: ``Data management is the development, execution
and supervision of plans, policies, programs and practices that control,
protect, deliver and enhance the value of data and information assets.''
While the term is somewhat clinical, does not necessarily capture the
essential creativity required (and is decidedly non-sexy), data
management may be the most appropriate phrase to describe the type of
data-related skills students need to make sense of the information
around them.

\subsubsection{SIDEBAR: Making bigger datasets accessible through
databases}\label{sidebar-making-bigger-datasets-accessible-through-databases}

Nolan and Temple Lang (2010) stress the importance of knowledge of
information technologies, along with the ability to work with large
datasets. Relational databases, first popularized in the 1970's, provide
fast and efficient access to terabyte-sized files. These systems use a
structured query language (SQL) to specify data operations. Surveys of
graduates from statistics programs have noted that familiarity with
databases and SQL would have been helpful as they moved to the
workforce.

Database systems have been highly optimized and tuned since they were
first invented. Connections between general purpose statistics packages
such as R and database systems can be facilitated through use of SQL.
Table 2 describes key operators for data manipulation in SQL.

\begin{verbatim}
verb          meaning
--------------------------------------------
SELECT      create a new result set from a table
FROM        specify table
WHERE       subset observations
GROUP BY    aggregate
ORDER       re-order the observations
DISTINCT    remove duplicate values
JOIN        merge two data objects
\end{verbatim}

Table 2: Key operators to support data management and manipulation in
SQL (structured query language)

Use of a SQL interface to large datasets is attractive as it allow the
exploration of datasets that would be impractical to analyze using
general purpose statistical packages. In this application, much of the
heavy lifting and data manipulation is done within the database system,
with the results made available within the general purpose statistics
package.

The ASA Data Expo 2009 website
(\url{http://stat-computing.org/dataexpo/2009}) provides full details
regarding how to download the Expo data (1.6 gigabytes compressed, 12
gigabytes uncompressed through 2008), set up a database using SQLite
(\url{http://www.sqlite.org}), add indexing, and then access it from
within R or RStudio. This is very straightforward to undertake (it took
the first author less than 2 hours to set up using several years of
data), though there are some limitations to the capabilities of SQLite.

MySQL (\url{http://www.mysql.com}, described as the world's most popular
open source database) and PostgreSQL are more fully-featured systems
(albeit with somewhat more complex installation and configuration).

The use of SQL within R (or other systems) is straightforward once the
database has been created (either locally or remotely). An add-on
package (such as \texttt{RMySQL} or \texttt{RSQLite}) must be installed
and loaded, then a connection made to a local or remote database. In
combination with tools such as R Markdown (which make it easy to provide
a template and support code, described in detail in ``Five Concrete
Reasons Your Students Should Be Learning to Analyze Data in the
Reproducible Paradigm'',
\url{http://chance.amstat.org/2014/09/reproducible-paradigm}) students
can start to tackle more interesting and meatier questions using larger
databases set up by their instructors. Instructors wanting to integrate
databases into their repertoire may prefer to start with SQLite, then
graduate to more sophisticated systems (which can be accessed remotely)
using MySQL.

The ${\tt dplyr}$ package encapsulates and replaces the SQL interface
for either system. It also features \emph{lazy} evaluation, where
operations are not undertaken until absolutely necessary.

Another option in \texttt{dplyr} is for the user to directly specify SQL
SELECT commands (this is an important topic for statistics majors to see
at some point in their programs). For example, the following code would
replicate the creation of the dataset of counts for the three airports
used to create Figure 2 using SQL (as opposed to using the interface
within ${\tt dplyr}$).

\begin{Shaded}
\begin{Highlighting}[]
\NormalTok{flights <-}\StringTok{ }
\StringTok{  }\KeywordTok{dbGetQuery}\NormalTok{(con, }\StringTok{"SELECT Dest, Year, Month, DayOfMonth, DayOfWeek, sum(1) as numFlights}
\StringTok{    FROM ontime WHERE (Dest = 'ALB' OR Dest = 'BDL' OR Dest = 'BTV') }
\StringTok{    GROUP BY Year, Month, DayOfMonth, Dest"}\NormalTok{)}
\end{Highlighting}
\end{Shaded}

In this example, a set of variables are selected (along with a derived
variable which sums the number of flights) from the three airports of
interest. The results are aggregated by day and destination (at which
point there are more manageable). The ${\tt dbGetQuery}$ function in the
${\tt RMySQL}$ package returns a dataframe containing the results from
the SQL SELECT call. The SQL syntax is similar, but not identical, to
the \texttt{dplyr} syntax.

Is setting up a database too much effort? We think not (and provide
further guidance at the aforementioned website). As another option,
those willing to explore can undertake similar analyses using the
${\tt nycflights13}$ package on CRAN, which includes five dataframes
that can be accessed within R (see the previous link for example files).

\end{document}